\documentclass[conference]{IEEEtran}
\usepackage{comment}

\usepackage{amssymb}
\usepackage{amsfonts}

\usepackage{subcaption}
\usepackage{graphicx}
\usepackage{xcolor}
\usepackage{multirow}
\usepackage{cite}
\usepackage{textcomp}
\usepackage{amsmath,graphicx}

\makeatletter
\newcommand{\figcaption}[1]{\def\@captype{figure}\caption{#1}}
\newcommand{\tblcaption}[1]{\def\@captype{table}\caption{#1}}

\def\vector#1{\mbox{\boldmath $#1$}} 
\makeatletter
\renewcommand\subsubsection{\@startsection{subsubsection}{3}{\z@}%
{0.5ex\@plus 0ex \@minus -.5ex}%
{0.5ex\@plus 0ex}
{\normalfont\normalsize\itshape}
}
\makeatother

\begin{document}
\title{Privacy-Preserving SVM Computing by Using Random Unitary Transformation}
\author{\IEEEauthorblockN{Takahiro Maekawa\IEEEauthorrefmark{1}, Takayuki NAKACHI\IEEEauthorrefmark{2}, Sayaka Shiota\IEEEauthorrefmark{1} and Hitoshi Kiya\IEEEauthorrefmark{1}}\\
\IEEEauthorblockA{\IEEEauthorrefmark{1}Tokyo Metropolitan
University, Tokyo, 191-0065, Japan\\maekawa-takahiro, sayaka@tmu.ac.jp, kiya@tmu.ac.jp\\\\
\IEEEauthorrefmark{2}NTT Network Innovation Laboratories, Kanagawa, 239-0847, Japan\\nakachi.takayuki@lab.ntt.co.jp}}

\maketitle

\begin{abstract}
A privacy-preserving Support Vector Machine (SVM) computing scheme is proposed in this paper. Cloud computing  has been spreading in many fields. However, the cloud computing has some serious issues for end users, such as unauthorized use and leak of data, and privacy compromise. We focus on templates protected by using a random unitary transformation, and consider some properties of the protected templates for secure SVM computing, where templates mean features extracted from data. The proposed scheme enables us not only to protect templates, but also to have the same performance as that of unprotected templates under some useful kernel functions. Moreover,  it can be directly carried out by using well-known SVM algorithms, without preparing any algorithms specialized for secure SVM computing. In the experiments, the proposed scheme is applied to a face-based authentication algorithm with SVM classifiers to confirm the effectiveness.
\end{abstract}
\IEEEpeerreviewmaketitle


\section{Introduction}
\label{sec:intro}
Cloud computing and edge computing have been spreading in many fields,
with the development of cloud services. However, the computing
environment has some serious issues for end users, such as
unauthorized use and leak of data, and privacy compromise, due to
unreliability of providers and some accidents.
While, a lot of studies on secure, efficient and flexible
communications, storage and computation have been reported \cite{Huang, Lazzeretti, Barni}.
For securing data, full encryption with provable security (like RSA,
AES, etc) is the most secure option. However, many multimedia
applications have been seeking a trade-off in security to enable other
requirements, e.g., low processing demands, retaining bitstream
compliance, and flexible processing in the encrypted domain, so that a
lot of perceptual encryption schemes have been studied as one of the
schemes for achieving a trade-off \cite{Lagendij, Ito1, Chuman2, Zhou, Kurihara_1, Kurihara, Chu_Kuri_1, Chu_Kuri_2, Chu_Iida_1, Chu_Kuri_3}

In the recent years, considerable efforts have been made in the
fields of  fully homomorphic encryption and multi-party
computation \cite{Araki1, Araki2, Lu, Toshinori}. However, these schemes can not be applied yet to SVM
algorithms, although it is possible to carry out some statistical
analysis of  categorical and ordinal data. Moreover, the schemes have
to prepare algorithms specialized for computing encrypted data.

Because of such a situation, we propose a privacy-preserving SVM
computing scheme in this paper . We focus on templates protected by
using
a random unitary transformation, which have been studied as one of
methods for cancelable biometrics \cite{Rathgeb, Nandakumar, Rane, Wright, Nakamura1, Nakamura2, Georghiades}, and then  consider some
properties of the protected templates for secure SVM computing, where
templates mean features extracted from data. As a result, the
proposed scheme enables us not only to protect templates, but also to
have the same performance as that of unprotected templates under some
useful kernel functions as isotropic stationary kernels.  Moreover, it can be directly carried out by
using well-known SVM algorithms, without preparing any algorithms
specialized for secure SVM computing. In the experiments, the proposed
scheme is applied to a face recognition algorithm with SVM classifiers
to confirm the effectiveness.

\section{preparation}
\subsection{Support Vector Machine}
Support Vector Machine (SVM) is a supervised machine learning algorithm which can be used for both classification or regression challenges, but it is mostly used in classification problems. In SVM, we input a feature vector \vector{x} to the discriminant function as
\begin{equation}
\label{eq:eq_sign}
\begin{split}
y = \mathrm{sign}(\vector{\omega}^T\vector{x}+b)\\
\lefteqn{\hspace{-60mm}with}\\
   \mathrm{sign}(u)=\begin{cases}
     1 & (u>1) \\
    -1 & (u\leq0)
  \end{cases},
  \end{split}
\end{equation} 
where $\vector{\omega}$ is a weight parameter, and  $b$ is a bias. 

SVM also has a technique called the kernel trick, which is a function that takes low dimensional input space and transform it to a higher dimensional space. These functions are called kernels. The kernel trick could be applied to Eq. (\ref{eq:eq_sign}) to map an input vector on further high dimension feature space, and then to linearly classify it on that space as  
\begin{equation}
\label{eq:eq_kernel_sign}
y = \mathrm{sign}(\vector{\omega}^T\phi(\vector{x})+b).
\end{equation}
The function $\phi(\vector{x}):\mathbb{R}^d\to\mathcal{F}$ maps an input vector $\vector{x}$ on high dimension feature space $\mathcal{F}$, where $d$ is the number of the dimensions of features. 
In this case, feature space $\mathcal{F}$ includes parameter $\vector{\omega}$ ($\vector{\omega}\in\mathcal{F}$).
The kernel function of two vectors $\vector{x}_i$, $\vector{x}_j$ is defined as
\begin{equation}
K(\vector{x}_i,\vector{x}_j)=\langle \phi(\vector{x}_i), \phi(\vector{x}_j)\rangle,
\end{equation}
where $\langle\cdot, \cdot\rangle$ is an inner product. There are various kernel functions. For example, Radial Basis Function(RBF) kernel is given by
\begin{equation}
\label{eq:rbf}
K(\vector{x}_i,\vector{x}_j)=\exp(-\varUpsilon \| \vector{x}_i - \vector{x}_j \|^2 )
\end{equation}
and polynomial kernel is provided by
\begin{equation}
K(\vector{x}_i,\vector{x}_j)=(1+\vector{x}_i^T\vector{x}_j)^l,
\end{equation}
where $\varUpsilon$ is a high parameter to decide the complexity of boundary determination, $l$ is a parameter to decide the degree of the polynomial, and $T$ indicates transpose.

This paper aims to propose a new framework to carry out SVM with protected vectors.

\subsection{Scenario}
\begin{figure}[t]
  \centering\includegraphics[width = 8cm]{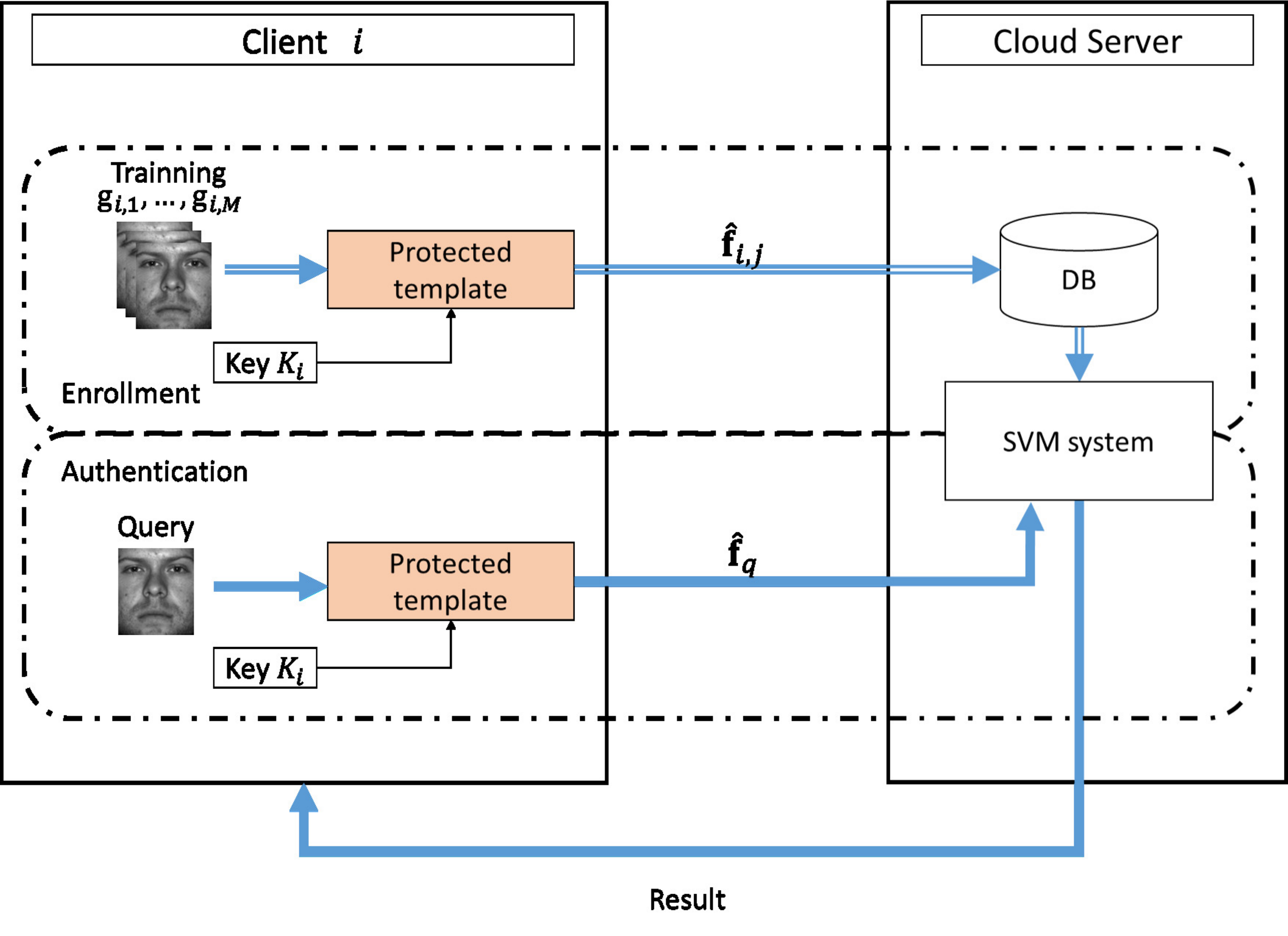}
 \caption{Scenario}
 \label{fig:architecture}
\end{figure}

Figure \ref{fig:architecture} illustrates the scenario used in this paper. In the enrollment, client $i$, $i \in \{1,2,...,N\}$, prepares training samples $\mathrm{g}_{i,j}, j \in \{1,2,...,M\}$ such as images, and a feature set ${\vector{\mathrm{f}}_{i,j}}$, called a template, is extracted from the samples. Next the client creates a protected template set $\hat{\vector{\mathrm{f}}}_{i,j}$ by a secret key $p_i$ and sends the set to a cloud server. The server stores it and implements learning with the protected templates for a classification problem. 

In the authentication, Client $i$ creates a protected template as a query and sends it to the server. The server carries out a classification problem with a learning model prepared in advance, and then returns the result to Client $i$. 

Note that the cloud server has no secret keys and the classification problem can be directly carried out by using well-known SVM algorithms.  In the other words, the server does not have to prepare any algorithms specialized for the classification in the encrypted domain.

\section{Proposed framework}
In this section, protected templates generated by using a random unitary matrix are conducted, and a SVM computation scheme with the protected templates is proposed under some kernel functions.
\subsection{Template Protection}

\label{sec:Uniprotect}
Template protection schemes based on unitary transformations have been studied as one of methods for cancelable biometrics\cite{Rathgeb, Wright, Nakamura1,Nakamura2, Nandakumar, Rane}. This paper has been inspired by those studies. 

A template $\vector{\mathrm{f}_{i,j}} \in \mathbb{R}^d$ is protected by a unitary matrix having randomness with a key $p_i$, $\vector{\mathrm{Q}}_{p_i} \in \mathbb{C}^{N\times N}$ as, 
\begin{equation}
\label{eq:trans}
\hat{\vector{\mathrm{f}}}_{i,j}=T(\vector{\mathrm{f}}_{i,j},{p_i})=\vector{\mathrm{Q}}_{p_i}\vector{\mathrm{f}}_{i,j},
\end{equation}
where $\hat{\vector{\mathrm{f}}}_{i,j}$ is the protected template. Various generation schemes of $\vector{\mathrm{Q}}_{p_i}$ have been studied to generate unitary or orthogonal random matrices such as Gram-Schmidt method, random permutation matrices and random phase matrices\cite{Nakamura2, Nakamura1}. For example, the Gram-Schmidt method can be  applied to a pseudo-random matrix to generate $\vector{\mathrm{Q}}_{p_i}$. Security analysis of the protection schemes have been also considered in terms of brute-force attacks, diversity and irreversibility.

\subsection{SVM with protected templates}
\label{sec:SVMpro}
\subsubsection{Properties}
Protected templates generated according to Eq. (\ref{eq:trans}) have the following properties under $p_i=p_s$\cite{Nakamura2}.

\noindent
\hspace{15pt}Property 1 : Conservation of the Euclidean distances:\\
          \begin{equation}\| \vector{\mathrm{f}}_{i,j} - \vector{\mathrm{f}}_{s,t} \|^2  =  \| \hat{\vector{\mathrm{f}}}_{i,j} - \hat{\vector{\mathrm{f}}}_{s,t} \|^2. \nonumber\end{equation} \\
\hspace{15pt}Property 2 : Conservation of inner products:\\
          \begin{equation}\langle \vector{\mathrm{f}}_{i,j},\vector{\mathrm{f}}_{s,t}\rangle=\langle\hat{\vector{\mathrm{f}}}_{i,j},\hat{\vector{\mathrm{f}}}_{s,t}\rangle, \nonumber\end{equation}\\
\hspace{15pt}Property 3 : Conservation of correlation coefficients: \\
           \begin{equation}\frac{\langle \vector{\mathrm{f}}_{i,j},\vector{\mathrm{f}}_{s,t}\rangle}{\sqrt{\langle \vector{\mathrm{f}}_{i,j},\vector{\mathrm{f}}_{s,t}\rangle}\sqrt{\langle \vector{\mathrm{f}}_{i,j},\vector{\mathrm{f}}_{s,t}\rangle}}=\frac{\langle\hat{\vector{\mathrm{f}}}_{i,j},\hat{\vector{\mathrm{f}}}_{s,t}\rangle}{\sqrt{\langle\hat{\vector{\mathrm{f}}}_{i,j},\hat{\vector{\mathrm{f}}}_{s,t}\rangle}\sqrt{\langle\hat{\vector{\mathrm{f}}}_{i,j},\hat{\vector{\mathrm{f}}}_{s,t}\rangle}}. \nonumber\end{equation}\\
where $\vector{\mathrm{f}}_{s,t}$ is a template of another client $s, s \in \{1,2,...,N\}$, who has M training samples $\mathrm{g}_{s,t}, t \in \{1,2,...,M\}$.

\subsubsection{Classes of kernels}
We consider applying the protected templates to a kernel function. In the case of using RBF kernel, the following relation is satisfied from property 1 and Eq.(\ref{eq:rbf})
\begin{eqnarray}
\label{eq:propKernel}
K(\hat{\vector{\mathrm{f}}}_{i,j},\hat{\vector{\mathrm{f}}}_{s,t})
&=&\exp(- \varUpsilon \| \hat{\vector{\mathrm{f}}}_{i,j} - \hat{\vector{\mathrm{f}}}_{s,t} \|^2 )\nonumber\\
&=&K(\vector{\mathrm{f}}_{i,j},\vector{\mathrm{f}}_{s,t})
\end{eqnarray}

A stationary kernel $K_S(\vector{x}_i - \vector{x}_j)$ is one which is translation invariant:
\begin{equation}
K(\vector{x}_i, \vector{x}_j) = K_S(\vector{x}_i - \vector{x}_j),
\end{equation}
that is, it depends only on the lag vector separating the two vectors $\vector{x}_i$ and $\vector{x}_j$. Moreover, when a stationary kernel depends only on the norm of the lag vectors between two vectors, the kernel $K_I( \| \vector{x}_i - \vector{x}_j \|)$ is said to be isotropic (or homogeneous)\cite{Genton}, and is thus only a function of distance:
\begin{equation}
K(\vector{x}_i, \vector{x}_j) = K_I( \| \vector{x}_i - \vector{x}_j \|).
\end{equation} 
For examples, RBF, WAVE and Rational quadratic kernels belong to this class, i.e, isotropic stationary kernel, called kernel class 1 in this paper. If kernels are isotropic, the propose scheme is useful under the kernels.

Besides, from property 3, we can also use a kernel $K_{In}( \langle\vector{x}_i, \vector{x}_j\rangle)$ that depends only on the inner products between two vectors given as
\begin{equation}
K(\vector{x}_i, \vector{x}_j) = K_{In}( \langle\vector{x}_i, \vector{x}_j\rangle).
\end{equation}
Polynomial kernel and linear kernel are in this class, referred to as class 2.

Some kernels such as Fisher and p-spectrum ones, to which the protected templates can not be applied, belong to other classes. We focus on using kernel class 1 and class 2.

\subsubsection{Dual problem}
Next, we consider binary classification that is the task of classifying the elements of a given set. A dual problem to implement a SVM classifier with protected templates is expressed as
\begin{equation}
\label{eq:eq_dual_kernel}
\begin{split}
&\max_\alpha\ \left(-\frac{1}{2}\sum_{\substack{i,s \in N\\j,t \in M}}\alpha_{i,j} \alpha_{s,t} y_{i,j} y_{s,t} \langle \phi(\hat{\vector{\mathrm{f}}}_{i,j}), \phi(\hat{\vector{\mathrm{f}}}_{s,t})\rangle + \sum_{\substack{i \in N\\j \in M}}\alpha_{i,j}
\right)\\
&s.t.\ \sum_{\substack{i \in N \\ j \in M}}\alpha_{i,j} y_{i,j} = 0, 0\leq\alpha_{i,j}\leq C,
\end{split}
\end{equation}
where $y_{i,j}$ and $y_{s,t}$$\in\{+1,-1\}$ are correct labels for each training data, $\alpha_{i,j}$ and $\alpha_{s,t}$ are dual variables and C is a regular coefficient.
If we use kernel class 1 or class 2 described above, the inner product $\langle\phi(\hat{\vector{\mathrm{f}}}_{i,j}), \phi(\hat{\vector{\mathrm{f}}}_{s,t})\rangle$ is equal to $K(\vector{\mathrm{f}}_{i,j},\vector{\mathrm{f}}_{s,t})$.
Therefore, 
even in the case of using protected templates, the dual problem with protected templates is reduced to the same problem as that of the original templates. This conclusion means that the use of the proposed templates gives no effect to the performance of the SVM classifier under kernel class 1 and class 2.

\subsection{Relation among keys}
\label{sec:key}
As shown in Fig \ref{fig:architecture}, a protected template $\hat{\vector{\mathrm{f}}}_{i,j}$ is generated from training data $\mathrm{g}_{i,j}$ by using a key $p_i$. Two relations among keys are summarized, here.
\subsubsection{Key condition 1: $p_1=p_2=...=p_N$}
The first key choice is to use a common key in all clients, namely, $p_1=p_2=...=p_N$. In this case, all protected templates satisfy the properties described in \ref{sec:SVMpro}, so the SVM classifier has the same performance as that of using the original templates. 
\subsubsection{Key condition 2: $p_1 \neq p_2 \neq .. .\neq p_N$}
The second key choice is to use a different key in each client, namely $p_1 \neq p_2 \neq .. .\neq p_N$. In this case, the three properties are satisfied only among templates with a common key. This key condition allows us to enhance the robustness of the security against various attacks as discussed later.

\section{Experimental Results}
The propose scheme was applied to face recognition experiments which were carried out as a dual problem.
\subsection{Data Set}
\begin{figure}[t]
\centering
  \begin{tabular}{c c c c}
     \begin{minipage}[b]{0.15\hsize}
 	 \centering\includegraphics[width = 1.5cm]{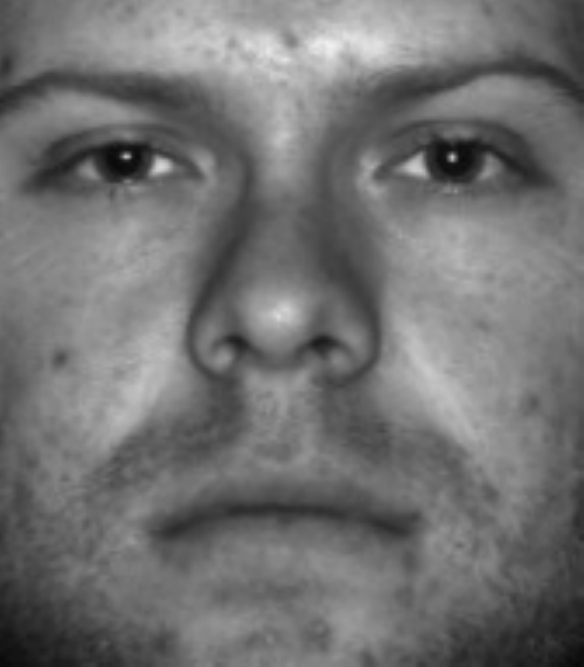} 
    \end{minipage}
    &
     \begin{minipage}[b]{0.15\hsize}
        \centering\includegraphics[width = 1.5cm]{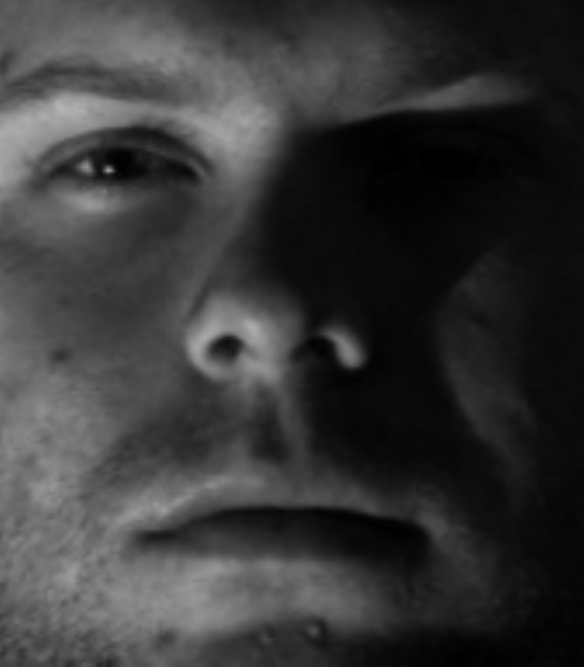}
      \end{minipage}
      &
       \begin{minipage}[b]{0.15\hsize}
 	 \centering\includegraphics[width = 1.5cm]{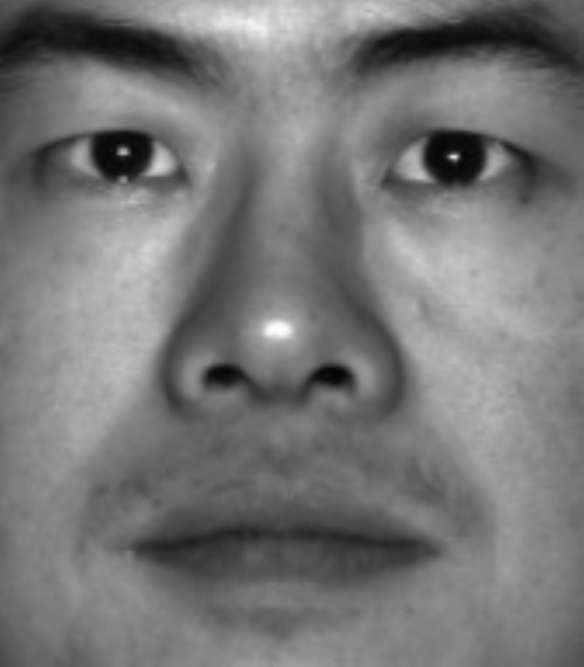}
    \end{minipage}    
    &
     \begin{minipage}[b]{0.15\hsize}
        \centering\includegraphics[width = 1.5cm]{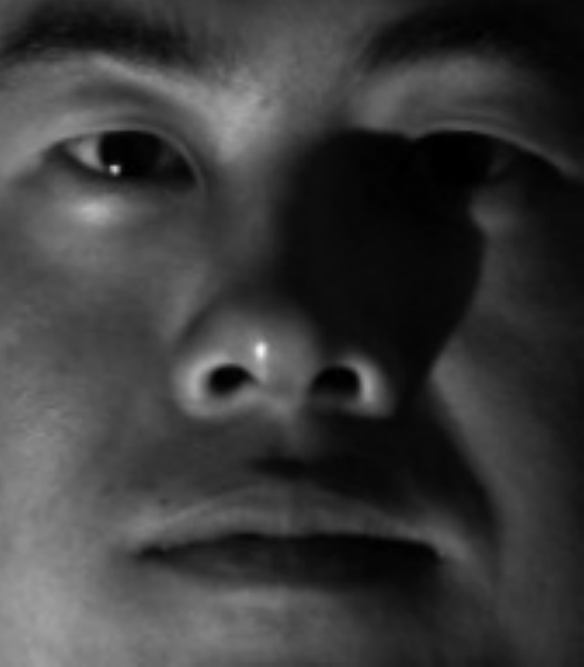}
    \end{minipage}   
      \\
\multicolumn{2}{c}{(a) person1} 
&
\multicolumn{2}{c}{(b) person2}
\end{tabular}
\caption{Examples of Extended Yale Face Database B}
\label{fig:db}
\end{figure}

\begin{figure}[t]
\centering
  \begin{tabular}{c c}
  \begin{minipage}[b]{0.2\hsize}
      \begin{center}
 	 \centering\includegraphics[width = 2cm]{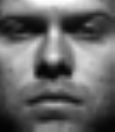}
      \end{center}
    \end{minipage}
    &
     \begin{minipage}[b]{0.2\hsize}
      \begin{center}
        \centering\includegraphics[width = 2cm]{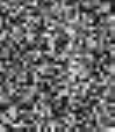}
      \end{center}
    \end{minipage}
  \\
   (a) template
   &
   (b) protected
\end{tabular}
\caption{An example of protection}
\label{fig:imageProtected}
\end{figure}

We used Extended Yale Face Database B\cite{Georghiades} that consists of 2432 frontal facial images with $192\times168$-pixels of $N=38$ persons like Fig \ref{fig:db}. 64 images for each person were divided into half randomly for training data samples and queries. We used random permutation matrices as unitary matrices to produce protected templates. Besides, RBF kernel and linear kernel were used, where they belong to kernel class 1 and class 2, respectively.
The protection was applied to templates with 1216 dimensions generated by the down-sampling method\cite{Wright}. The down-sampling method divides an image into non-overlapped blocks and then calculates the mean value in each block. Figure \ref{fig:imageProtected} shows the examples of an original template and the protected one.

\subsection{Results and Discussion}
In face recognition with SVM classifiers, one classifier is created for each enrollee. The classifier outputs a predicted class label and a classification score for each query template $\hat{\vector{\mathrm{f}}}_{q}$, where $\hat{\vector{\mathrm{f}}}_{q}$ is a protected template generated from the template of a query, $\vector{\mathrm{f}}_{q}$. The classification score is the distance from the query to the boundary ranging. The relation between the classification score $S_q$ and a threshold $\tau$ for the positive label of $\vector{\mathrm{f}}_{q}$ is given as
\begin{equation}
if \ S_q \geq \tau \ then\  accept;\  else\ reject.
\end{equation}
In the experiment, False Reject Rate(FRR), False Accept Rate(FAR), and Equal Error Rate(EER) at which FAR is equal to FRR were used to evaluate the performance.
\subsubsection{$p_1=p_2=...=p_N$}\par
\begin{figure}[t]
\centering\subcaptionbox{Linear kernel ($C=1$)\label{fig:fig1}}
{\includegraphics[width=7.5cm]{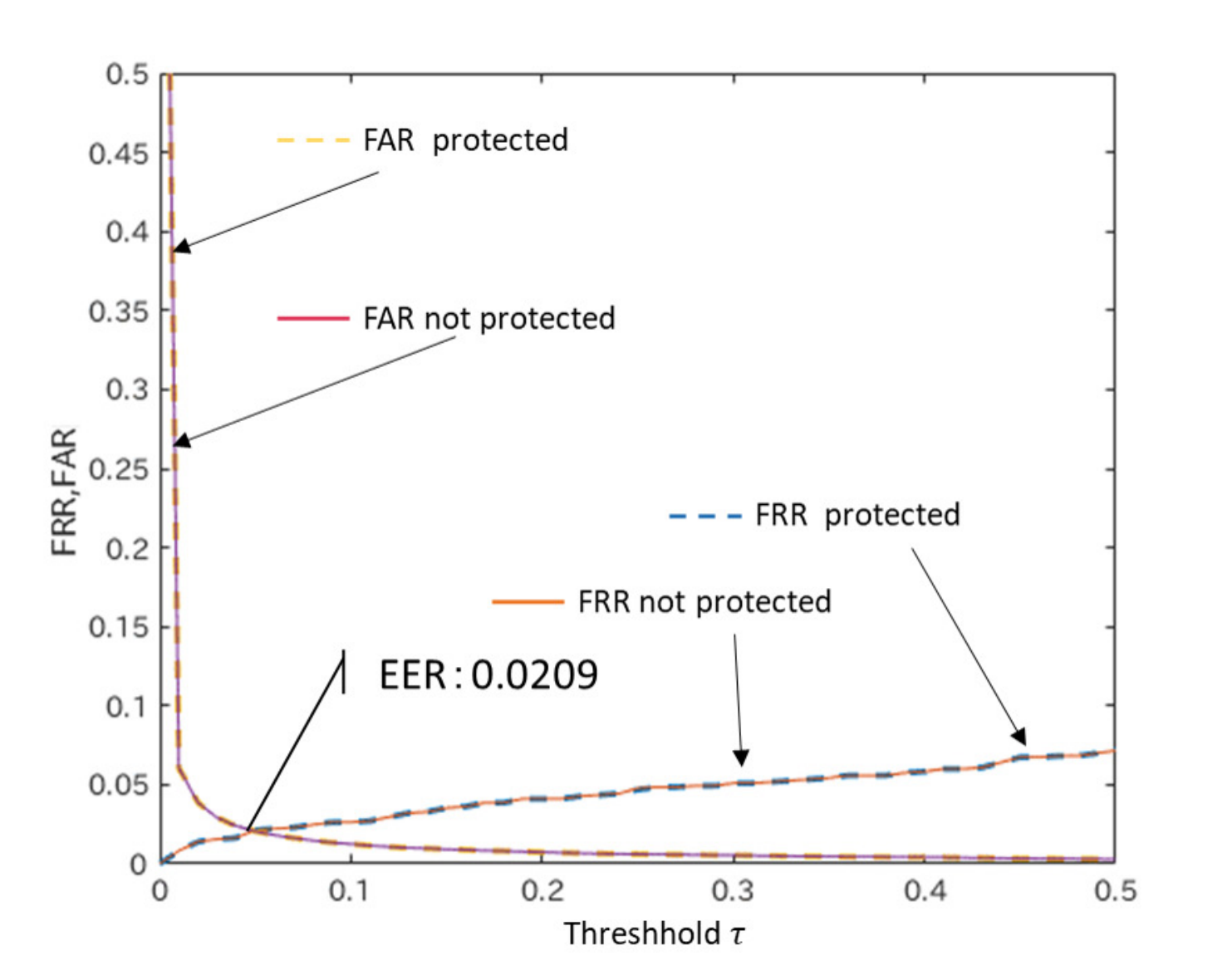}}

\centering

\centering\subcaptionbox{RBF kernel ($C=34$, $\varUpsilon=81$)\label{fig:fig1}}
{\includegraphics[width=7.5cm]{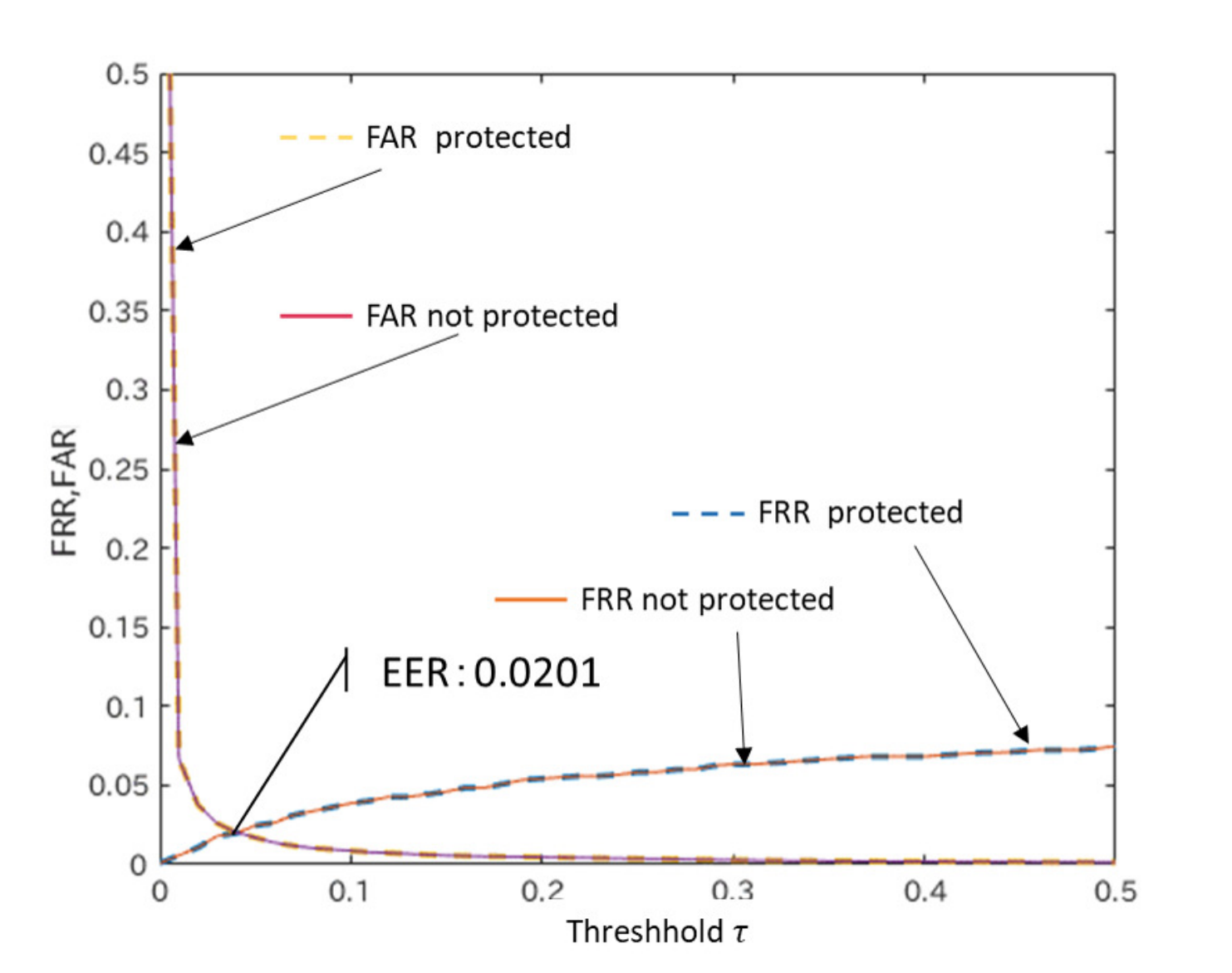}}
\caption{FAR and FFR ($p_1=p_2=...=p_N$)}\label{fig:result_1}
\end{figure}

Figure \ref{fig:result_1} shows results in the case of using key condition 1. The results demonstrate that SVM classifiers with protected templates (protected in Fig \ref{fig:result_1}) had the same performances as those fo SVM classifiers with the original templates (not protected in Fig \ref{fig:result_1}). From the results, it is confirmed that the proposed framework gives no effect to the performance of SVM classifiers under key condition 1.
\subsubsection{$p_1 \neq p_2 \neq .. .\neq p_N$}\par
Figure \ref{fig:result_2} shows results in the case of using key condition 2. In this condition, it is expected that a query will be authenticated only when it meets two requirements, i.e. the same key and the same person, although only the same person is required  under key condition1. Therefore, the performances in Fig. \ref{fig:result_2} were slightly different from those in Fig. \ref{fig:result_1}, so the FAR performances for key condition 2 were better due to the strict requirements.

\subsubsection{Unauthorized outflow ($p_1 \neq p_2 \neq .. .\neq p_N$)}\par
\begin{figure}[t]
  \centering\includegraphics[width = 7.5cm]{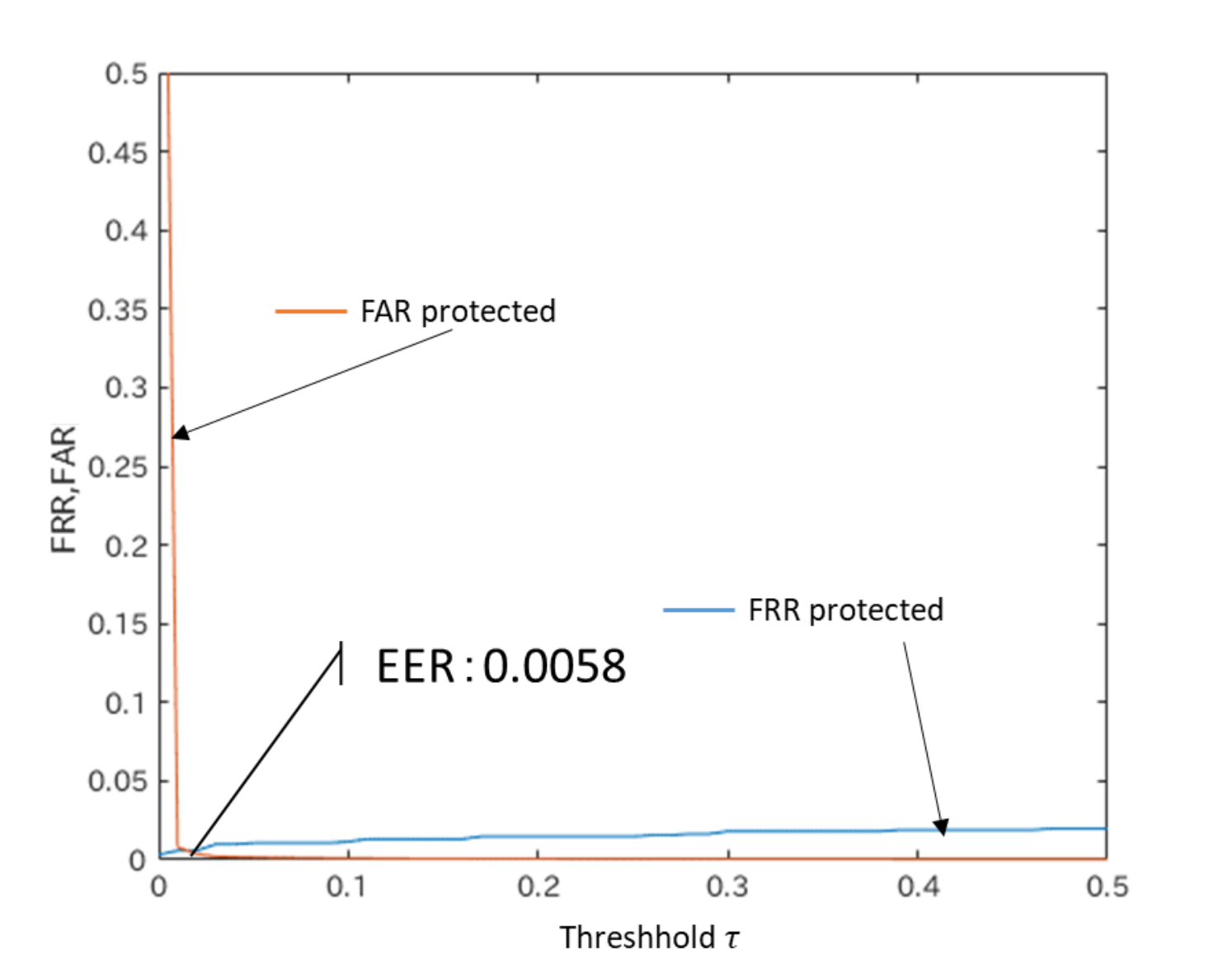}
 \caption{FAR and FAR (RBF kernel, $p_1 \neq p_2 \neq .. .\neq p_N$)}
 \label{fig:result_2}
\end{figure}
\begin{figure}[t]
  \centering\includegraphics[width = 7.5cm]{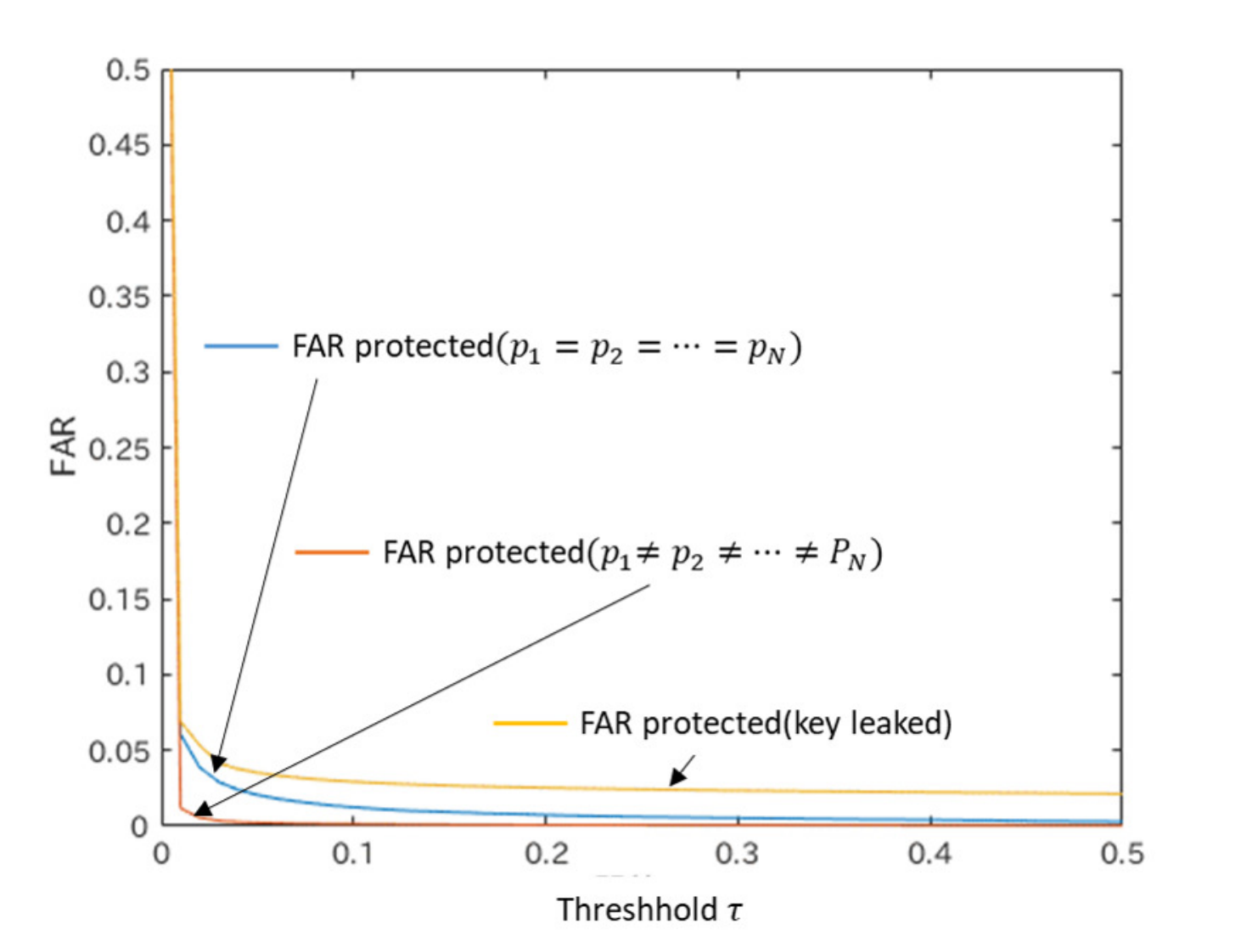}
 \caption{FAR with leaked keys (RBF kernel)}
 \label{fig:key_leak}
\end{figure}

Figure \ref{fig:key_leak} shows the FAR performance in the case that a key $p_i$ leaks out. In this situation, other clients could use the key $p_i$ without any authorization as spoofing attacks. As shown in Fig.\ref{fig:key_leak}, the FAR (key leaked in Fig.\ref{fig:key_leak}) still had low vales due to two requirements, although it was slightly degraded, compared to Fig.\ref{fig:result_2}. 
 
\begin{figure}[t]
  \centering\includegraphics[width = 7.5cm]{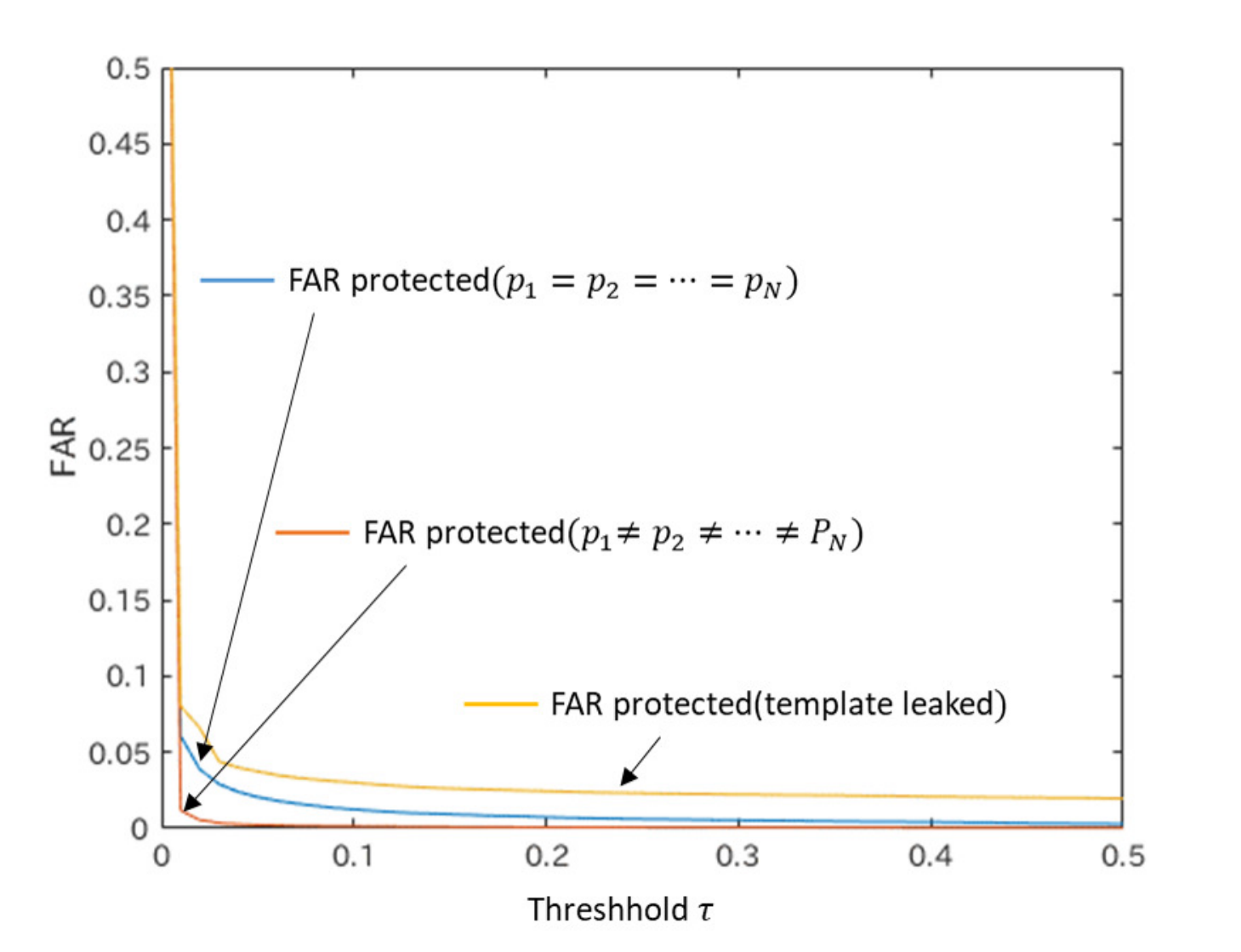}
 \caption{FAR with leaked original templates (RBF kernel)}
 \label{fig:template_leak}
\end{figure}

Figure \ref{fig:template_leak} is the FAR performance in the case that a template $\vector{\mathrm{f}}_{i,j}$ leaks out. It is confirmed that the FAR (template leaked in Fig.\ref{fig:template_leak}) still had low vales as well as in Fig.\ref{fig:key_leak}.

From these results, the use of key condition 2 enhances the robustness of the security against spoofing attacks.

\section{conclusion}
In this paper, we proposed a privacy-preserving SVM computing scheme with
protected templates. It was shown that templates protected by a
unitary transform has some useful properties,  and  the properties
allow us to securely compute SVM algorithms without any degradation of
the performances. Besides, two key conditions were considered to
enhance the robustness of the security against various attacks.  Some
face-based authentication experiments using SVM classifiers were
also demonstrated to experimentally confirm the effectiveness of the
proposed framework.

\subsection*{Acknowledgements}
This work was partially supported by Grant-in-Aid for Scientific
Research(B), No.17H03267, from the Japan Society
for the Promotion Science.
\\
\\
\\
\\
\\
\\
\\
\\
\\
\\
\\



\end{document}